\documentclass[format=acmsmall,authordraft=true]{acmart}

\usepackage{booktabs} 
\usepackage{array,multirow,graphicx}
\usepackage{hhline}
\usepackage{csquotes}
\usepackage{soul}
\usepackage{balance}
\usepackage[T1]{fontenc}

\usepackage{color}

\setcopyright{rightsretained}

\acmConference[TOIS]{ACM TOIS}{...}{} 
\acmYear{2021}
\copyrightyear{2021}

\acmPrice{15.00}

\begin{document}


\title{Feature Modulation to Improve Struggle Detection in Web Search: A Psychological Approach}

\author{Jiyun Luo}  
\affiliation{%
  \institution{Pinterest}
  \country{USA}}
\email{...}

\author{Yan	Yang}  \authornote{Work completed during the author worked as a research assistant at Georgetown.} 
\affiliation{%
  \institution{University of Nevada, Reno}
  \country{USA}}
\email{yy490@georgetown.edu}

\author{Valerie	Nayak} \authornote{Work completed during the author worked as an intern at Georgetown.} 
\affiliation{%
  \institution{Carnegie Mellon University}
  \country{USA}}
\email{vjn@andrew.cmu.edu}

\author{Grace Hui Yang} \authornote{Grace Hui Yang is the corresponding author.} 
\affiliation{%
  \institution{InfoSense, Department of Computer Science, Georgetown University}
  \streetaddress{37 and O Streets Northwest}
  \city{Washington}
  \state{DC}
  \postcode{20057}
  \country{USA}}
\email{huiyang@cs.georgetown.edu}

\begin{abstract}

Searcher struggle is important feedback to Web search engines. Existing Web search struggle detection methods rely on effort-based features to identify the struggling moments. Their underlying assumption is that the more effort a user spends, the more struggling the user may be. However, recent studies have suggested this simple association might be incorrect. This paper proposes a new feature modulation method for struggle detection and refers to the reversal theory in psychology. The reversal theory (RT) points out that instead of having a static personality trait, people constantly switch between opposite psychological states, complicating the relationship between the efforts they spend and the level of frustration they feel. Supported by the theory, our method modulates the effort-based features based on RT's bi-modal arousal model. Evaluations on week-long Web search logs confirm that the proposed method can statistically significantly improve state-of-the-art struggle detection methods. 

\end{abstract}

\settopmatter{printfolios=true}

\maketitle

\section{Introduction}
\label{sec_intro}


Searcher struggle is the event that a searcher ``makes strenuous efforts in the face of difficulties"\footnote{\url{merriam-webster.com}} during a search process. At a struggling moment, the user experiences a negative emotion of feeling  frustrated, upset, or annoyed by the search activity. Detecting searcher struggle is an important task for Web search engines because it can alert the search engines to timely adjust their algorithms. Most existing struggle detection methods are classifiers using features that measure user efforts~\cite{Li:2009:GAM:1571941.1571951,Jiang:2015:UPG:2684822.2685319}. This is because struggling behaviors are often demonstrated through excessive repetitions of user actions. For instance, when searching for ``curly hair dye", a user issued over twenty queries and still could not find a satisfactory result. It is straightforward to assume that a large number of user efforts suggest user struggles.

However, this conjecture does not always align with real-world observations. One observation is that non-struggling search tasks may also possess many user-issued queries and clicks. For instance, a user searched for ``guy proposing ideas," and the log shows that he browsed over 900 search results and bookmarked 27; the user reported he enjoyed the search throughout the session. 
Recent lab studies have raised similar concerns. Edwards and Kelly \cite{Edwards:2017:EFD:3077136.3080818} pointed out that although an increase in user effort might help predict searcher struggle, such an increase can also signify engagement or exploration. 
Without an effective mechanism, the distinguishing power of effort-based features diminishes, making it difficult to separate a negative experience such as struggle from positive experiences such as engagement and thus challenging to improve Web user experiences.

This paper refers to the psychology literature and proposes a new feature modulation method for struggle detection. Our work is motivated by the Reversal Theory (RT)~\cite{gerrig2015psychology,apter2001motivation,personalitydynamics,APTER1984265,fourdomains}. RT is a ``mode-based" psychology theory that challenges some fundamental assumptions in the fields of motivation and personality. Rather than assuming static personality traits, RT focuses on the ``complexity, changeability, and dynamics of human motivation and personality"~\cite{APTER1984265}. Its key insight is that people change in the flow of everyday life, and their personalities and motivations can reverse~\cite{personalitydynamics}. 
RT suggests some counter-intuitive ideas, such as whether a user is conforming to rules or rebelling has little impact on his struggling behaviors, but whether she works on a serious or playful task would matter. We conducted statistical hypothesis tests on these arguments and confirmed them on commercial search logs. 

Moreover, RT states that human motivations can be organized into a few dimensions 
and each dimension consists of two opposing states. Within one dimension, a person can reverse between the pair of states but can be only active at one of them at any given moment,  which indeed suggests a bi-modal distribution instead of a single-modal distribution. We leverage this insight in this paper and propose a novel and effective feature modulation method for struggle detection. We propose to (1) select features responsible for the related RT dimension, (2) moderate the selected features by reducing the bias between the two distributions in the bi-modal arousal model, and (3) send the modulated features to classifiers to identify if a search session has struggling moments. 

Our method can be used in combination with any feature-based struggle detection method. We evaluate our approach on one-week-long search logs collected from both mobile and PC platforms. 
The experimental results show that our method is highly effective; it can significantly improve a few top-performing methods by $\sim$5\% accuracy and $\sim$9\% positive precision.



\section{Related Work}
\label{sec_related}



\subsection{Struggle Detection}

Studies on searcher struggles can be grouped into (1) laboratory studies  and (2) query log studies. Both types of studies look for meaningful relationships between searcher struggles and their search behaviors. 

{\it Lab studies} monitor a user's entire search process in a laboratory setting and collect explicit user feedback via questionnaires. They ask a user if they are experiencing a struggling moment during a search session and study interesting behavior patterns when the struggle happens. For instance, Aula et al. \cite{Aula:2010:SBC:1753326.1753333} found that when encountering a struggle, a user tends to (a) formulate question-like queries, (b) use advanced search operators, (c) spend more time examining search results, (d) be more likely to write the most extended query in the middle of a search session if the search eventually fails and (e) at the end of the session if the search succeeds. They also suggested that task difficulty may lead to user struggles. Xu et al.'s lab study~\cite{10.1145/3331184.3331353} suggested that searcher struggles are related to the user's mood. When users are irritated or excited, they tend to issue more queries than in neutral moods. 
However, the work did not distinguish between negative emotions and positive emotions, leading to more queries. 
A highly relevant lab study to ours is Edwards and Kelly's work~\cite{Edwards:2017:EFD:3077136.3080818}. They observed that although user efforts increase might help predict searcher struggles, such increases can also indicate engagement, the opposite of struggles. These lab studies can go to great lengths to investigate searcher struggles; however, constrained by monetary costs, they usually only perform with small groups of users and the limited number of search tasks.

{\it Log-based studies} record a user's search process in search logs and analyze the historical data to understand the searcher's behaviors and how they relate to struggles. Usually, a struggling event is labeled afterward by third-party annotators. Log-based studies can be large-scale and support automatic detection of searcher struggles. Most methods derive helpful features from the logs and use regressors or classifiers to detect the struggles. 
For instance, Hassan et al. \cite{Hassan:2014:SED:2556195.2556221}  worked on detecting ``struggling'' and ``exploring'' (including ``exploring and struggling") search sessions. Their effort-based features included the number of unique queries, term additions, removals and substitutions, clicks, and dwell time. They reported accuracy of $81.67\%$ for detecting ``struggling'' sessions. They also acknowledged that a user behaves similarly when exploring and struggling; the search logs for both types of sessions are ``similar in terms of the number of queries and the session duration''. 
However, their focus was on finding new features that can help distinguish the subtle difference between exploring and struggling sessions, while ours is on new ways to re-use existing features.  
J. Li et al.~\cite{Li:2009:GAM:1571941.1571951} studied good abandonment, which is relevant to the absence of searcher struggles. Good abandonment happens when a user abandons her search before clicking any results as the content on the SERP has met the information need. When good abandonment happens, a user's effort is minimal, and struggle is absent. They also reported the important role of search topics in determining good abandonment, which is  investigated in our paper as significant features in the ``means-ends" motivational dimension. Feild et al. \cite{Feild:2010:PSF:1835449.1835458} compared features derived from  query logs and physical sensors. They found that using log-generated features is reliable and more effective than using sensor-generated features in detecting searcher struggles. 

Our work belongs to the log-based studies. Although we use many prior features~\cite{Hassan:2014:SED:2556195.2556221,Odijk:2015:SSW:2806416.2806488}, our work is a novel method to modulate these features for more effective struggle detection.

Other well-studied, negative search experiences besides struggles include irrelevancy and dissatisfaction~\cite{Feild:2010:PSF:1835449.1835458}. 
Note that these concepts are related to struggles but not interchangeable. For instance, dissatisfaction occurs after a search task when a user has not found satisfactory information from the search results. On the other hand, struggles can occur anytime during a session, as soon as the results are frustrating. 
Even if a user is satisfied at the end of a session, she may still experience struggles during it. Our paper only studies struggles. 




\subsection{Reversal Theory} 

Reversal theory (RT) is a psychology theory on personality dynamics and motivations. It recognizes that people ``are essentially changeable and move between different motivational styles" \cite{apter2001motivation}. This theory ``sheds light on the paradoxes of risk-taking, addiction, rebelliousness, and other areas of motivation, emotion, and personality"~\cite{apter2001motivation}. 

The key ideas in RT are the following. (1) In everyday life, people's motivations can be organized along a few
dimensions. They include ``means-ends", ``rules", ``transactions", and ``relationships". (2) Each dimension consists of a pair of opposing states.\footnote{Some books call these states ``meta-motivational states", ``motivational styles," or motives. For simplicity, we call them motivational states or states in this paper.} 
(3) A person can only be at one of the two states at any given moment. (4) A person can reverse between the pair of motivational states. (5) Although each person has their ``dominating" states, i.e., they have a preference to stay more often in a state when in a non-dominant state, people follow the current state to the same extent as they are at the dominating state. 

Reversal theory ``challenges some of the basic concepts of mainstream psychology, such as the trait concept of personality'' \cite{apter2001motivation}. For example, RT ``explains how anxiety can be reverted almost instantaneously into excitement and vice versa" \cite{michael1989reversal}. It also explains that a person ``may now experience his job as an obligation and at a later time experience the very same job as a kind of game." In another example, RT improves the {\it model of arousal}~\cite{HebbArousal} in traditional motivational theories from a single-modal model into a bi-modal model, which measures the complex relationship between one's happiness level\footnote{Also called hedonic level in some books.} and effort level.\footnote{Also known as arousal level.} This new bi-modal model of arousal is the basis for our work to perform feature modulation. 

The first two RT dimensions, ``means-ends" and ``rules," describe how users perform tasks. They are relevant to our discussion and examined in this paper. The last two dimensions, ``transactions" and ``relationships," describe interpersonal interactions instead of user and task; thus are less relevant and not discussed in this paper.

\section{Problem Formulation}
\label{background}

\subsection{Web Search Struggle Detection}

\begin{table*}[!ht]
 \centering \footnotesize
 \caption{Effort-Based Features. * marks new features. ** marks feature groups selected for feature modulation (Section 5.2). Other features' references are provided.} 
 \label{tab-features}
 \begin{tabular}{|p{0.48\linewidth}|p{0.48\linewidth}|}
 \hline
 \multicolumn{1}{|c|}{\textbf{Efforts to Query**}} & \multicolumn{1}{|c|}{\textbf{Efforts to Scroll}}  \\
  \cline{1-2}
 Number of queries in a session \cite{Edwards:2017:EFD:3077136.3080818} &  Screen size* \\
 Number of unique queries in a session \cite{Edwards:2017:EFD:3077136.3080818} & Total and avg. number of scrolling down actions* \\
 \cline{2-2}
 Avg. number of terms per query \cite{Edwards:2017:EFD:3077136.3080818}  & \multicolumn{1}{c|}{\textbf{Efforts to Re-formulate Queries}}  \\
 \cline{2-2}
   
 Avg. number of characters per query \cite{Edwards:2017:EFD:3077136.3080818} & Avg. cosine similarity between every query and the first query \cite{Hassan:2014:SED:2556195.2556221} \\

Percentage of manually-typed queries \cite{Edwards:2017:EFD:3077136.3080818} &  Avg. cosine similarity of every query pair in a session \cite{Hassan:2014:SED:2556195.2556221} \\

Percentage of suggested queries (that are automatically corrected, suggested, or completed by the search engine)  \cite{Edwards:2017:EFD:3077136.3080818} &  Avg. edit distance per adjacent query pair \cite{Hassan:2014:SED:2556195.2556221}   \\

  The longest query's position in a session*  &  Number of query generations (when removing one or more terms from its previous query)\cite{Hassan:2014:SED:2556195.2556221} \\
 
 \cline{1-1}
 \multicolumn{1}{|c|}{\textbf{Efforts to Click**}} & Number of query specifications (one or more terms are added into its previous \\
 \cline{1-1}   
 
Total and avg. number of clicks in \cite{Feild:2010:PSF:1835449.1835458, Hassan:2014:SED:2556195.2556221} & query~\cite{Hassan:2014:SED:2556195.2556221} \\
 
Total  and avg. number of Satisfactory (SAT) clicks \cite{Feild:2010:PSF:1835449.1835458, Hassan:2014:SED:2556195.2556221}  & Difference between the first query length and the avg. query length* \\

Percentage of queries without clicks \cite{Fox:2005:EIM:1059981.1059982, Kim:2014:MDT:2556195.2556220, Huffman:2007:WRR:1277741.1277839} & Standard deviation of query lengths in a session*  \\ 

Maximum and avg. number of adjacent queries without clicks*  & Avg. number of terms appear in the previous query \cite{Hassan:2014:SED:2556195.2556221}  \\

Total and avg. number of images clicked in a session*   &  Avg. number of terms added to the previous query \cite{Hassan:2014:SED:2556195.2556221}  \\

Total and avg. number of ads clicked in a session*    & Avg. number of terms deleted from the previous query \cite{Hassan:2014:SED:2556195.2556221}  \\
  
Total and avg. number of bookmarks clicked in a session*   & Avg. number of terms that substitute terms in the previous query \cite{Hassan:2014:SED:2556195.2556221} \\
   
\cline{2-2} 
Number of events (clicks, bookmarks, and queries) in a session*  & \multicolumn{1}{c|}{\textbf{Efforts to Diversify**}} \\
\cline{2-2}

Number of clicks at the first two queries* & Percentage of unique URLs among all clicked URLs \cite{Hassan:2014:SED:2556195.2556221}   \\

Number of clicks at the third and fourth queries*   & Percentage of the unique domain (DNS) names  
among all clicked URLs \cite{Hassan:2014:SED:2556195.2556221}   \\

Number of clicks at the fifth and sixth queries* & Total number of unique clicks*  \\

Whether the session ends with a click* &  Total number of unique topics  \cite{Hassan:2014:SED:2556195.2556221}   \\

\cline{1-1}  
\multicolumn{1}{|c|}{\textbf{Efforts to Read**}}  &  Entropy of topic distribution in a session \cite{Hassan:2014:SED:2556195.2556221}  \\
\cline{1-1}  

\cline{2-2} 
Total dwell time of all clicks \cite{Feild:2010:PSF:1835449.1835458, Hassan:2014:SED:2556195.2556221}  &  \multicolumn{1}{c|}{\textbf{Efforts to Issue Rare Queries  \& Rare Clicks}}  \\
\cline{2-2} 

Avg. number of image impressions per SERP*  &   Log(1 + avg. query frequency in popularity data) \cite{Huffman:2007:WRR:1277741.1277839, Odijk:2015:SSW:2806416.2806488, Hassan:2014:SED:2556195.2556221} \\

Total number of zoom-in on result images* & Log(1 + a query's avg. SAT clicks in popularity data) \cite{Huffman:2007:WRR:1277741.1277839, Odijk:2015:SSW:2806416.2806488, Hassan:2014:SED:2556195.2556221}  \\
 
Log(1 + avg. dwell time per click in a session)* & Log(1 + a query's avg. clicks in popularity data) \cite{Huffman:2007:WRR:1277741.1277839, Odijk:2015:SSW:2806416.2806488, Hassan:2014:SED:2556195.2556221} \\
 
Log(1 + avg. dwell time per click exclude clicks for the last query)* &   A query's avg. click entropy in the popularity data  \cite{Huffman:2007:WRR:1277741.1277839, Odijk:2015:SSW:2806416.2806488, Hassan:2014:SED:2556195.2556221}  \\


Log(1 + time passed until the first SAT click)*  & Log(1 + a query's avg. number of fast-back clicks (whose dwell time is less \\

Log(1 + avg. time spent on each SERP in a session)*  & than 15s) in the popularity data) \cite{Huffman:2007:WRR:1277741.1277839, Odijk:2015:SSW:2806416.2806488, Hassan:2014:SED:2556195.2556221}   \\

Log(1 + avg. time spent on each SERP exclude the last query)* &  Log(1 +  a clicked URL's avg. click frequency in the popularity data)* \\
 \hline
 \end{tabular}
 \end{table*}



A Web search {\bf struggle} is an event that a user feels frustrated by the search results at some point during a search session. 
We formulate the task of Web search struggle detection as a binary classification problem, with the two classes ``struggling" and ``non-struggling". The time unit to study whether a struggle happen is chosen to be a single search session, as it is 1) a natural block for a search task and 2) produces more stable responses than every single user action. Here we define a {\bf struggling session} as a session that contains a struggling moment, where a user feels frustrated by the search results at some point during the session. A {\bf non-struggling session} is a session that contains no struggling moment throughout the session. In this work, we obtain search sessions by segmenting them from query logs (See details in Section~\ref{experimentsetup}).

 
We denote a struggle predictor $Y$ and an input search session $s$, described by a vector of features $X(s)$. The probability of $s$ being struggling is $P(Y = 1|X(s), \Theta)$, with $\Theta$ being the model parameter. The classifier predicts ``struggling'' when $P(Y = 1 |X(s), \Theta) > 0.5$ and ``non-struggling'' otherwise. In Section~\ref{sec_experiment}, we experiment on multiple classifiers to show the effect of the proposed feature modulation method. The class labels are obtained by third-party manual annotation (Section~\ref{experimentsetup}).

\subsection{Features} 

The input feature vector $X(s)$ is automatically extracted from a query log. Our features include features proposed by prior works and new features presented in this paper. Most features are indicators of certain user efforts, which measure the quantity and diversity of user actions in a session.
Table \ref{tab-features} lists these effort-based features and groups them into seven feature groups. 

{\it Efforts to Query.} The first group of features measures user efforts spent on writing queries.
We derive most of them from Edwards and Kelly's work  \cite{Edwards:2017:EFD:3077136.3080818}. Usually, many queries in a session indicate a user has spent a lot of effort typing queries. But, redundant queries may be ``copy-and-pasted'', which requires little effort. 
Likewise, using system-suggested queries also requires little user effort. We, therefore, measure their ratios among all queries in the session. 
The new feature we propose is the longest query\textquotesingle s position in a session, inspired by Aula et al.'s work \cite{Aula:2010:SBC:1753326.1753333}, where they pointed out that the longest query often appears at the end of a successful session. 

{\it Efforts to Click.} The second group of features measures user efforts spent on clicking the search result URLs. 
Existing features include total numbers of clicks and satisfactory (SAT) clicks, widely used to infer relevance, satisfaction~ \cite{Fox:2005:EIM:1059981.1059982, Kim:2014:MDT:2556195.2556220, Huffman:2007:WRR:1277741.1277839}, as well as struggles~\cite{Feild:2010:PSF:1835449.1835458, Hassan:2014:SED:2556195.2556221}. In addition to them, we propose new features to measure clicks for different types of search results, including images, ads, and Web pages. 
We also propose new features indicating abandonment and little effort -- for instance, the average number of adjacent queries that receive  no clicks. 
Moreover, we also introduce bookmarked results and clicks on different phases of a session. 

{\it Efforts to Read.} The third feature group measures user efforts spent on reading and examining the content of search results \cite{Feild:2010:PSF:1835449.1835458, Hassan:2014:SED:2556195.2556221}. Except for the standard dwell time feature, we propose to count the number of \textit{zoom-in} actions on image results and specific types of \textit{dwell time} on different kinds of returning items and search results sections. 

{\it Efforts to Scroll.} The fourth feature group measures user efforts spent on reaching out to results beyond the current sight. These features are all new. Here we propose to count the numbers of scrolling-downs and screen-resizing. When a user scrolls down or resizes her screen, a new search request is sent to the back-end engine to get  and fresher search results for the same query. We obtain the numbers of scrolling-downs and resizing  by counting the number of pagination requests from the user. 

{\ Efforts to Reformulate Queries.} The fifth feature group measures user efforts spent on editing queries and re-articulating the information needs. 
\cite{Hassan:2014:SED:2556195.2556221}. 
We add new features to measure the variance of the query lengths after editing. 

{\it Efforts to Diversify.} The sixth feature group measures user efforts in diversifying the search results' content and the examination process. Features in this group indicate how exploratory the user and the search are. We mainly use features proposed by \cite{Hassan:2014:SED:2556195.2556221} for click diversity and topical diversity. A new feature added is the total number of unique clicks. 

{\it Efforts to Issue Rare Query \& Rare Clicks.} The seventh group of features measures user efforts spent on critical thinking and being novel and unique. They include rare queries and rare clicks that a user would create in a session compared to the large Web population who have the exact or similar information need. The idea is that issuing popular queries, like most others, requires fewer efforts, while giving a rare query requires more ``thinking" efforts. Likewise, clicking on unpopular URLs is also an indicator of critical thinking. We obtain the Web population's click data from a commercial search engine from 11/15/2020 to 11/21/2020 and use that as the basis to derive which queries and clicks are rare. 

Besides these seven feature groups, we also recruit features that are not effort-related. For instance, we use the taxonomy topic of the search task as a categorical feature. However, our method mainly acts on the effort-based features shown in Table \ref{tab-features}.

\section{The Reversal Theory} 

\begin{table}[t]
\centering\small
\caption{Motivation Dimensions and States.}
\label{tab-reversal-theory}
\begin{tabular}{ p{5cm}|p{5cm}}
\hline
\multicolumn{2}{c}{\textbf{Means-ends}}\\
\hline
Telic & Paratelic \\
$\bullet$ \textit{\small{Serious. Focus on future goals and achievement. Tend to avoid arousal, risk \& anxiety.}} & $\bullet$ \textit{\small{Playful, passion and fun. Focus on current moment. Seek excitement and entertainment.}} \\
\hline
\multicolumn{2}{c}{\textbf{Rules}}\\
\hline
Conformist & Negativistic \\
$\bullet$ \textit{\small{Conforming. Value rules and tradition. Tend to operate within rules and expectations.}} & $\bullet$ \textit{\small{Rebellious. Value innovation and changes. Like to explore new possibilities.}} \\
\hline
\multicolumn{2}{c}{\textbf{Transaction}}\\
\hline
Mastery & Sympathy \\
$\bullet$ \textit{\small{One wants to be in
control, whether this be over people, tasks, ideas, machinery or anything else that one can interact with.}} & $\bullet$ \textit{\small{Wanting to develop close and nurturing relationships, to be
tender and sensitive.}} \\ 
\hline
\multicolumn{2}{c}{\textbf{Relationships}}\\
\hline
Autic & Alloic \\
$\bullet$ \textit{\small{Doing things
for self rather than for others.}} & $\bullet$ \textit{\small{Genuinely concerned with
others, and putting them first.}} \\
\hline
\end{tabular} 
\end{table}

RT is a psychology theory that studies {\em personality dynamics}~\cite{personalitydynamics}. It states that 
in the flow of daily life, a person regularly reverses, like a teeter-totter, between opposing motivational states. This section introduces the basic concepts in RT and how they relate to Web search struggle detection. 

\subsection{Opposing Motivational States} 

RT groups human motivations into four dimensions (also known as ``domains"). They are ``means-ends", ``rules", ``transactions", and ``relationships" \cite{michael1989reversal}. The first two dimensions describe how a user performs tasks and are relevant to  this paper. The last two dimensions describe interpersonal interactions and are less relevant. Table \ref{tab-reversal-theory} shows the first two dimensions in RT. 

The first dimension, ``means-ends", is about achievements, goals, and enjoyments of process. It has two opposing motivational states that ``reflect people's motivational styles, the meaning they attach to a given situation at a given time, and the emotion they experience"~\cite{personalitydynamics}. The two states are  {\it telic}, at which one is motivated by achievement, task completion, and fulfilling of goals; and {\it paratelic}, at which one is playful and seeks excitement and fun. When one is at the telic state, she is serious about the task at hand and focuses on achieving the task goal, whereas when one is paratelic, the activity she does is not for the sake of the task's goal but the task's own sake. E.g., people run because they enjoy running themselves, not because they want to win a medal~\cite{personalitydynamics}. 

In the context of web search, telic and paratelic states can parallel to goal-oriented and non-goal-oriented search tasks. A user at the telic state would focus on finishing the search task, for instance, to look for job vacancies or medical help. A user at the paratelic state would seek to enjoy the search process itself, for example, browsing for fun videos on YouTube. 


The second dimension, ``rules", is how routines, expectations, and constraints could direct a person's activity. The two opposing states are  {\it conformist}, at which a person tends to operate within rules and expectations; and {\it negativistic}, at which a person wishes to push against regulations and explore new possibilities. An example for a conformist is ``I am eating because this is what I am supposed to do at this moment."~\cite{personalitydynamics}. And when one thinks ``I am eating because I am not supposed to eat at this moment", she is at the negativistic state. Interestingly, the same behavior can be motivated by opposite reasons. 

In the context of web search, conformist and negativistic are parallel to non-exploratory and exploratory search behaviors. When a user is in a conformist state, she obeys rules and meets others' expectations. For instance, the user would use the query suggested by the search engine instead of creating her own. On the contrary, when a user is in a negativistic state, she can explore  novel ideas. For instance, she would issue rare queries, read from different URLs, and prefer more novel and diverse search results. 



\subsection{RT's Bi-Modal Arousal Model}

\begin{figure}[t]
    \centering
        \includegraphics[width=0.6\textwidth]{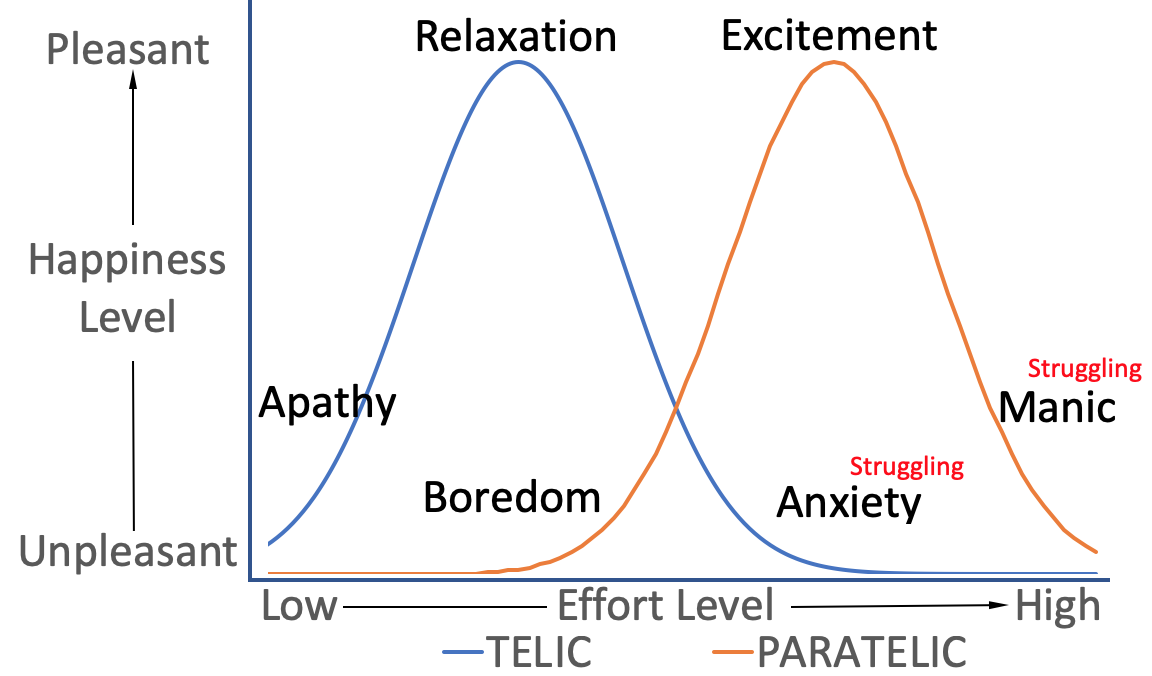}
   \caption{Arousal Model for Means-ends; Both Anxiety and Manic indicate struggling.~(adapted from~\cite{reversal}).} 
    \label{fig:twocurves} 
\end{figure}

In psychology, the {\it model of arousal}~\cite{HebbArousal} tells the relationship between a person's happiness level and arousal level. Arousal is a central concept in motivational theories, which indicates the intensity of activities and feelings a person experiences. In our context, it corresponds to the intensity of search actions, i.e., the level of effort. Therefore, the arousal model can help establish a relationship between a user's happiness and effort levels. 

The traditional model of arousal is a single-modality model. It suggests that as the arousal level increases, a single optimal arousal level exists to reach the happiest moment~\cite{HebbArousal}. For instance, there is an optimal usage level of air-conditioning to feel the most comfortable; too much or too little would both reduce a person's happiness. It suggests an inverted U shape or a {\it Gaussian} distribution. However, this model cannot capture extreme happiness caused by  intense arousal, e.g., riding a roller-coaster. It can neither capture that people experience a high level of happiness with low arousal, e.g., being calm and happy after completing a significant project. 

On the contrary, RT's arousal model is a bi-modality model. RT assumes two optimums are present. Each of these is for one of the two opposite states within a motivational dimension. The model takes the shape of two inverted U-curves or two Gaussian distributions crossing. Figure \ref{fig:twocurves}~\cite{reversal} illustrates this bi-modal arousal model for the means-ends dimension. 
Here the x-axis is effort, and the y-axis is happiness. A low happiness level indicates negative feelings. Among the negative emotions there are apathy, boredom, anxiety, and manic. Both  ``anxiety" and ``manic" happen when efforts are substantial, and happiness is low. In this paper, we consider  both of them are struggling and do not distinguish them further. On the graph, the two curves each represent one of the two states, telic or paratelic. We can see the two states peak at different effort levels -- The telic curve peaks early when a moderate amount of effort happens;  While the paratelic curve peaks late after a significant amount of effort are present. 

In the context of Web search, this bi-modal arousal model could be the cause of inconsistent prediction of user struggles because the same level of user efforts can indeed map to two different happiness levels, depending on at which state the user is at the moment. 
For instance, the same effort level could mean ``struggling/anxiety" for a user at the telic state and ``excitement" for a user at the paratelic state. To resolve this inconsistency, we propose to shift the two (state) curves horizontally closer to each other until they overlap and then separate the left and right end, given that the struggling instances lie at the right end. 





\subsection {``Rules" Dimension is Irrelevant} 
\label{sub_sec_rule_dimension}

\begin{figure}[t]
    \centering
        \includegraphics[width=0.8\linewidth]{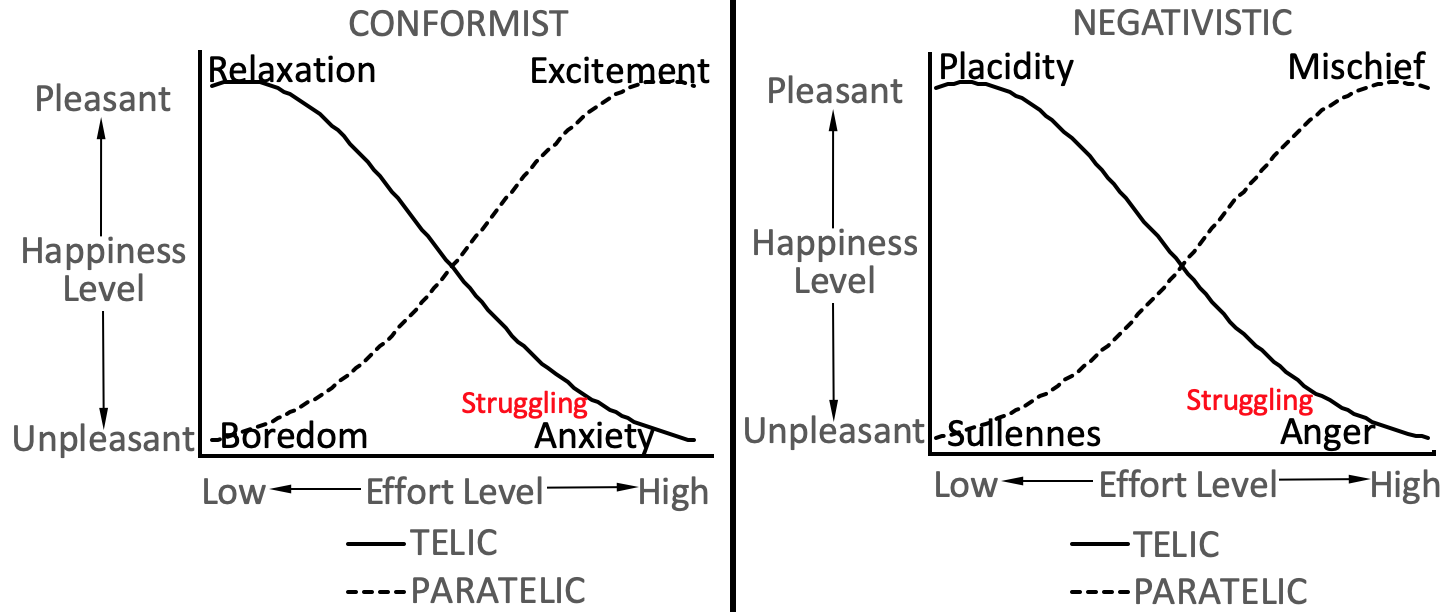}
   \caption{Interplay of ``means-ends" and ``rules" (adapted from~\cite{apter2001motivation}).}
    \label{fig:4curves} 
\end{figure}


As we mentioned before, the first two RT dimensions are seemingly relevant to Web search because they care about users and tasks. However, contrary to our intuition, RT suggests that the ``rules" dimension has little impact on struggling, and only the ``means-ends" dimension matters. It \cite{apter2001motivation}'s interplay of the first two dimensions (Figure \ref{fig:4curves}).  The two sub-figures in Figure \ref{fig:4curves} depicts RT's arousal model when the second dimension state is ``conformist" and ``negativistic," respectively. We notice that in both sub-figures, struggles happen at the same effort level, which suggests that whether the user is ``conformist" or ``negativistic" has little impact on determining struggles. 

\begin{figure}[t]
    \centering
       \includegraphics[width=0.45\linewidth]{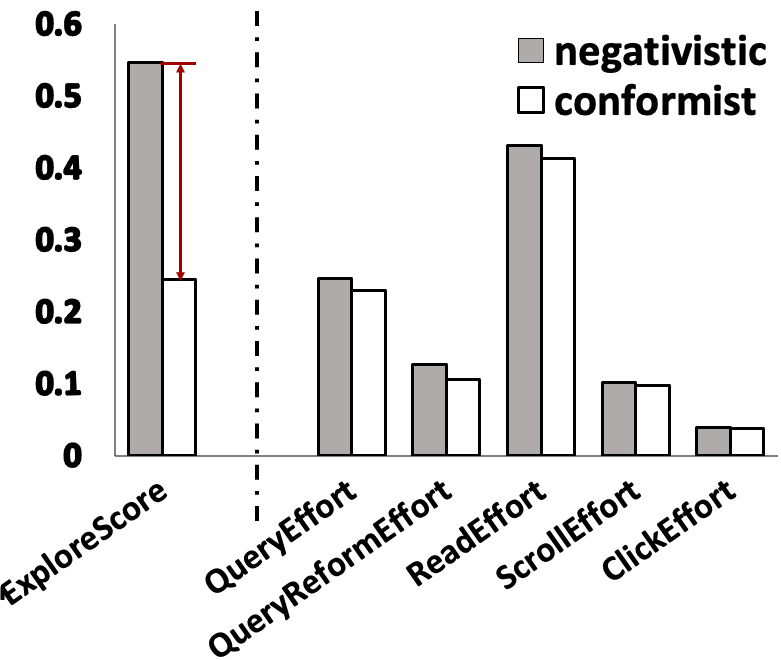}
     \caption{Differences in Features at Two ``Rules"  States. {\color{red}$\updownarrow$} marks significant differences at p$<$.05.} 
     \label{fig-task-user-bias2} 
\end{figure}

To confirm this, we conducted two MANOVA hypothesis tests, one for the first RT dimension and one for the second, on a whole week's query log collected from a commercial search engine (more details about the data in Section \ref{sec_experiment}). For the ''rules" dimension,  we make the following hypotheses:

$H_0$: The ``rules" dimension is irrelevant to a user's happy level. In other words, there is no statistically significant difference in the average effort level from users at the conformist state and users at the negativistic state.

$H_1$: The ``rules" dimension is relevant to a user's happy level. The average effort spent by users in the conformist state differs from that spent in the negativistic state.

We carry out the hypothesis test in the following steps. First,  we sort all search sessions in the query log-based on an $ExploreScore$. We define the \textit{ExploreScore}; it is the average score of features in the ``efforts to diversity" and ``efforts to issue rare queries and clicks" feature groups:
\begin{equation}
 \begin{aligned}
 ExploreScore =   \frac{1}{|F_{diverse}|} \sum_{i\in F_{diverse}} f_i + \frac{1}{|F_{rare}|} \sum_{j\in F_{rare}} f_j
 \end{aligned}
 \end{equation}
where $f_i$ is a feature in the feature group $F_{diverse}$ and $f_j$ is a feature in group $F_{rare}$. All features are normalized into $[0,1]$ before taking the average. A bigger {\it ExploreScore} suggests a more ``negativistic" state, where a user puts more effort in diversifying the search process and being against conventions. A smaller {\it ExploreScore} suggests a more ``conformist" state, where the user puts less effort in doing so. 

Second, we establish the ``conformist" and ``negativistic" states from the query log data. To do so, we select the top 15\% (we empirically choose 15\% to relax a bit from a rigorous top 10\%) sessions with the highest $ExploreScore$ to represent the negativistic state and the last 15\% sessions to illustrate the conformist state. 


Third, we conduct a statistical significance test between the two states for all feature groups except the two groups used to calculate $ExploreScore$. For each remaining feature group, we obtain the ``state averages" for features in the group at the two states. Then we 
conduct a MANOVA \cite{french2008multivariate} test across all feature groups and 5 ANOVA \cite{STHLE1989259} tests for each of them. The detailed results are: \textit{MANOVA} [F(5, 330)=1.1352, p=0.3414], \textit{QueryEffort} [F(1, 334)=0.5753, p=0.4487], \textit{QueryReformEffort} [F(1, 334)=1.5423, p=0.2151], \textit{ReadEffort} [F(1, 334)=1.2738, p=0.2599], \textit{ScrollEffort} [F(1, 334)=1.6459, p=0.2004], and \textit{ClickEffort} [F(1, 334)=0.5863, p=0.4444]. 
The significance tests produce $p>.05$ and fail to reject the null hypothesis. In other words, the ``rules" dimension is irrelevant to a user's happiness level, which implies it is irrelevant to struggle detection and confirms what is suggested by RT. 


Further, we plot the mean feature values for the conformist and negativistic states in Figure \ref{fig-task-user-bias2}. We can see that, except for the feature groups used to generate $ExploreScore$, none of the other feature groups show a statistically significant difference between the two states. Again, this confirms RT suggests that when the first two dimensions interplay, the ``rules" dimension has little impact on user efforts and struggle detection. We, therefore, do not handle features along this dimension. 

A similar MANOVA hypothesis test runs for the ``means-ends" dimension. That result is statistically significant and confirms what is suggested by RT that the first ``means-ends" dimension is influential to a user's struggle. We, therefore, use ``means-to-ends" as the primary dimension for our research.

\section{Our Approach}

\begin{figure}[t]
    \centering
       \includegraphics[width=0.45\linewidth]{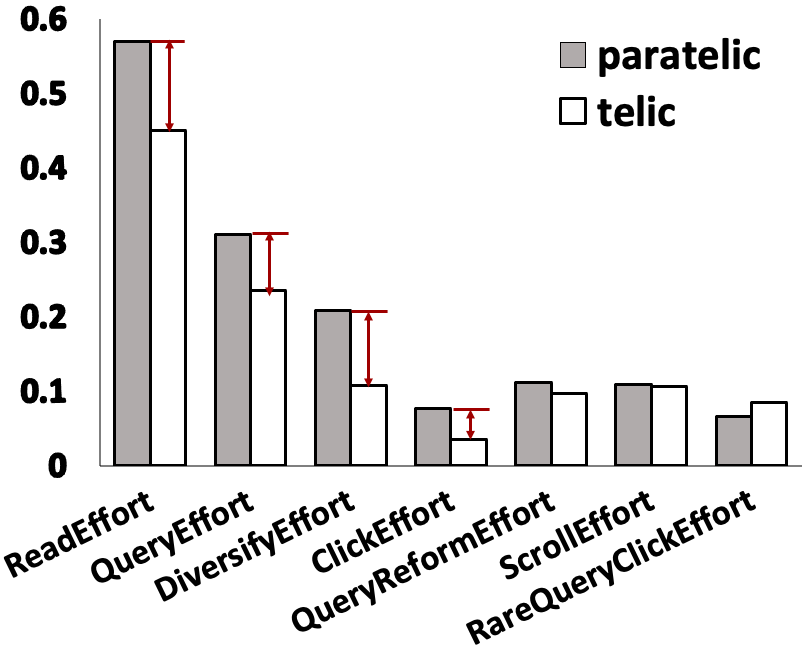}
   \caption{Feature value gaps along ``Means-ends". The red {\color{red}$\updownarrow$} indicates the difference is statistically significant at p$<$.05.}   
    \label{fig-task-user-bias1}
\end{figure}

This paper proposes a novel feature modulation method for struggle detection based on RT's bi-modal arousal model. First, we establish the two ``means-ends" motivational states, telic and paratelic, for every search session. Based on RT, a search session would be at any one moment only at either state, not both. Second, based on what RT's interplay figure suggests and our hypothesis tests confirm, we select highly related features to the ``means-ends" dimension. 
Third, we modulate these features by shifting their values for those in the paratelic state towards those in the telic state until their arousal model's peaks overlap. Fourth, we use the modulated features to fit a classifier and predict whether a session has struggles. 



\subsection{Put Sessions into ``Means-Ends" States}
\label{means_ends_dimen}

RT's bi-modal model of arousal (Figure \ref{fig:twocurves}) tells us that without knowing which motivational state the user is in, it is challenging to separate struggling from excitement or boredom from relaxation. We are thus motivated to (1) detect which state the user (and the session) is at, and then (2) move the two curves closer to each other for a selected group of features so that the struggles would be separable from the rest. Figure \ref{fig-modulation} illustrates our idea. 

Our first step is to put every session into either a telic or paratelic state. The bi-modal arousal model is a two-component Gaussian mixture model, whose means and variances can be found by the Expectation-Maximization (EM) algorithm \cite{EM}. In the mixture model, a data point can have a soft mapping onto both Gaussians. However, based on RT, at any one moment, a user can only be at one of the opposing states, not both. We choose to follow what RT suggests in this work and only associate a search session with one of the two states. We, therefore, propose to take a less common approach to identify the states for each session. 
We propose to assign the sessions into states based on the session's topic. RT considers telic states are associated with ``serious" tasks, and paratelic states are associated with ``playful" tasks~\cite{personalitydynamics}. Other research also pointed out that search topic shows the impact on searcher behaviors~\cite{Li:2009:GAM:1571941.1571951}.  We determine a session's search topic using a taxonomy. Without losing generality, an internal taxonomy, constructed by graph-based algorithms ~\cite{10.1145/3366423.3380271, 10.1007/978-3-030-30796-7_26}, is used. Note that the proposed method is general and should be compatible with most other Web taxonomies. To determine a session's search topic, first, we extract every clicked URL in the session. Second, we assign each clicked URL to a taxonomy category. It measures the similarity between the URL link's text with the category's name using tf-idf and word2vec cosine similarity. The taxonomy category with the highest similarity score to the URL text becomes the label to the URL. We use Gradient Boosted Decision Trees (GBDT) to combine the similarity scores (with a learning rate of 0.1, minimum split loss 0.5, and maximum tree depth 8). Third, We chose the most frequent URL label in the session  as the search topic for the session. 

We assign those with a search topic relating to serious, significant events, such as financial, health, and career decisions, to a telic state. For instance, ``Health", ``Job," and ``Finance." To a paratelic state, we assign those with a search topic relating to fun, relaxing events, such as entertainment and hobby. For instance,  ``Entertainment", ``Art'', and ``Beauty''.



\subsection{Select ``Means-Ends" Features}

RT suggests that we should modulate the features along the ``means-ends" dimension only. We therefore select features that are significant along the means-ends dimension, aka being able to distinguish the two ``means-ends" states. Other feature groups would remain the same without modulation. 

First, we normalize all effort-based features within a feature group into the range $[0,1]$ using $\frac{value - min\_value}{max\_value - min\_value}$. Second, we calculate two ``state average" scores for each feature group by taking the group average for sessions at the telic and paratelic states. Third, we conducted a MANOVA test to compare the state average score for all feature groups in the two states. The significance test result [F(7, 292)=34.3121, p$<$0.0001] proves that these feature groups are statistically significantly affected by the two states, which agrees with what RT suggests. Fourth, we then conducted one ANOVA test for each feature group to select the significant features. We find that four out of seven features groups,  \textit{ReadEffort} [F(1, 298)=39.4581, p$<$0.0001],  \textit{QueryEffort} [F(1,298)=95.3286, p$<$0.0001], \textit{DiversifyEffort} [F(1,298)=30.0176, p$<$0.0001], and  \textit{ClickEffort} [F(1,298)=54.4846, p$<$0.0001], are statistically significantly different in paratelic and telic sessions. 

Figure \ref{fig-task-user-bias1} plots the mean feature values from each selected feature group. As we can see, the four feature chosen groups show a large gap between the telic and paratelic sessions.  We determine these feature groups as ``means-ends" features and modulate them. 




\subsection{Modulate the Features}


\begin{figure}[t]
\includegraphics[width=0.9\linewidth]{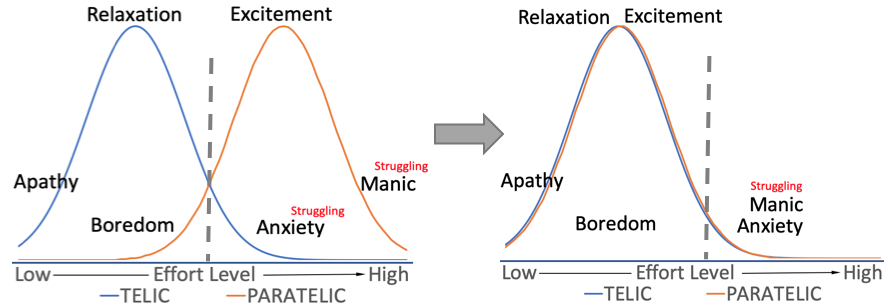}
  \caption{Modulation to separate struggles from non-struggles. }
  \label{fig-modulation} 
\end{figure}



Although we did not use the EM algorithm to find the means and variances of the two Gaussian distributions, our state assignment method based on the search topic still roughly forms Gaussian distributions, as what RT states. We, therefore, leverage this information to remove the bias between the two Gaussians. 

Given a feature $X_i$ $\in$ \{\textit{QueryEffort}, \textit{ClickEffort}, \textit{ReadEffort}, \textit{DiversifyEffort}\}, we use $X^{i}_{telic}$ and $X^{i}_{paratelic}$ to represent two different Gaussian distributions, each for $X_i$'s feature values in the telic state and paratelic state, respectively: 
\[X^{i}_{telic} \sim \mathcal{N}(\mu_{i_{telic}}, \sigma_{i_{telic}}^{2}).\]
and 
\[X^{i}_{paratelic} \sim \mathcal{N}(\mu_{i_{paratelic}}, \sigma_{i_{paratelic}}^{2}).\] 
where $\mu_{i_{telic}}$ and $\sigma_{i_{telic}}$ are the mean and standard deviation of the $i^{th}$ feature in all telic sessions; and $\mu_{i_{paratelic}}$ and $\sigma_{i_{paratelic}}$ are the mean and standard deviation of the $i^{th}$ feature in all paratelic sessions. We obtain the states as described in Section 5.1 and calculate the means and variances directly from them. 

We propose to reduce the bias between the two distributions by a Bayesian scaling method, shifting the paratelic towards the telic state for the selected ``means-ends" features. This transformation is done by
\begin{equation}
\label{eq-modulation}
    X_{_{paratelic}}' = \frac{\sigma_{_{telic}}}{\sigma_{_{paratelic}}}X_{_{paratelic}} + \mu_{_{telic}} - \frac{\sigma_{_{telic}}}{\sigma_{_{paratelic}}}\mu_{_{paratelic}}
\end{equation}
where $X_{_{paratelic}}$ is the original feature value in the paratelic state, and $X_{_{paratelic}}'$ is the new feature value after modulation. 

As illustrated in Figure \ref{fig-modulation},  the effort levels previously identified as both ``anxiety/struggling"  and ``excitement" would now be separable after feature modulation. We can now combine these modulated ``means-ends" features and other un-modulated features with a wide range of classifiers for struggle detection.

\section{Experimental Results}
\label{sec_experiment}

We conduct experiments to evaluate our method. 
The design of the experiments focuses on showing the before-and-after effect of using feature modulation in struggle detection.

\begin{table}[t]
\centering\small
\caption{Dataset statistics.}
\label{tab-dataset_statistic}
\begin{tabular}{| p{0.8cm}|p{1cm}|p{1cm}|p{1cm}|p{1.6cm}|p{1.8cm} |}
\hline
& Duration & \#Sessions & \#Struggle & \#NonStruggle & \#Query/Session \\
\hline
Mobile & Nov.  & 	1,123 & 299 &	824 & 5.39 \\
\cline{1-1} \cline{3-6}
PC & 22 $\sim$ 28, 	     &	1,034	 &  302 & 	732 & 4.62 \\
\cline{1-1} \cline{3-6}
Total & 2020 & 2,157 & 601 & 1,556 & 5.02 \\ 
\hline 
\end{tabular}
\end{table}

\begin{table*}[t]
\centering\footnotesize
\caption{Mobile: performance of struggle detection (Up and down arrows indicate absolute performance increase and decrease. $\dagger$ shows statistically significant improvement from ``feature modulation (X+FM)'' runs over the original runs (one-tailed t-test, p$=$.01)). ``X+FMNS'' refers to ``Feature Modulation with No feature Selection''.} 
\label{tab-result-1}
\begin{tabular}{|p{2.3cm}|p{0.7cm}|p{0.9cm}|p{0.7cm}|p{0.9cm}|p{0.7cm}|p{0.9cm}|p{0.7cm}|p{0.9cm}|p{0.7cm}|p{0.9cm}|}
\hline
 & accu. & impr. & pos. p & impr. & pos. r & impr. &  neg. p & impr. & neg. r & impr. \\
\hline
\hline
ZeroRule & 0.7337 & --  & -- & --  & 0.0000 & --  & 0.7337 & --  & 1.0000 & --  \\
\hline
\hline
LM &	0.8413 & &	0.7239 & &	0.6561 & &	0.8788 & &	0.9088 & \\
LM+FMNS & 0.8621 & 2.5\%$\uparrow$ & 0.7342 & 1.4\%$\uparrow$ & 0.6393 & 2.5\%$\downarrow$ & 0.8946 & 1.8\%$\uparrow$ & 0.9297 & 2.3\%$\uparrow$\\
LM+FM &	0.8910 & 5.9\%$\uparrow\dagger$ &	0.7513 & 3.7\%$\uparrow\dagger$ &	0.6116 & 6.8\%$\downarrow$ & 0.9157 & 4.2\%$\uparrow\dagger$ & 0.9542 & 5.0\%$\uparrow\dagger$\\
\hline
SVM & 0.8565 & & 0.7956 & & 0.7414 & &	0.8823  & &	0.9105 & \\
SVM+FMNS & 0.8675 & 1.3\%$\uparrow$ & 0.7940 & 0.2\%$\downarrow$ & 0.7180 & 2.9\%$\downarrow$ & 0.8929 & 1.2\%$\uparrow$ & 0.9260 & 1.7\%$\uparrow$\\
SVM+FM & 0.8928 & 4.2\%$\uparrow\dagger$ &	0.8511 & 7.0\%$\uparrow\dagger$ &	0.7278 & 1.8\%$\downarrow$ & 0.9052 & 2.6\%$\uparrow$ & 0.9533 & 4.7\%$\uparrow\dagger$\\
\hline
\hline
Hassan et al.\cite{Hassan:2014:SED:2556195.2556221}  & 0.8507 & & 0.7729 & &	0.6439 & &	0.8737 & & 0.9287 & \\
Hassan et al.+FMNS & 0.8626 & 1.4\%$\uparrow$ & 0.7962 & 3.0\%$\uparrow$ & 0.6447 & 1.3\%$\uparrow$ & 0.8807 & 0.8\%$\uparrow$ & 0.9408 & 1.3\%$\uparrow$ \\
Hassan et al.+FM & 0.8786 & 3.3\%$\uparrow\dagger$ &	0.8419 & 8.9\%$\uparrow\dagger$ & 0.6923 & 7.5\%$\uparrow\dagger$ &	0.8894 & 1.8\%$\uparrow$ &	0.9501 & 2.3\%$\uparrow$\\
\hline
MART & 0.8740 & &	0.7968 & &	0.7305 & &	0.9002 & & 0.9288 &\\
MART+FMNS & 0.8835 & 1.1\%$\uparrow$ & 0.8042 & 0.9\%$\uparrow$ & 0.7144 & 2.2\%$\downarrow$ & 0.9065 & 0.7\%$\uparrow$ & 0.9409 & 1.3\%$\uparrow$ \\
MART+FM &  0.9055 & 3.6\%$\uparrow\dagger$ & 0.8666 & 8.8\%$\uparrow\dagger$ & 0.7754 & 6.1\%$\uparrow\dagger$ &	0.9182 & 2.0\%$\uparrow$ & 0.9548 & 2.8\%$\uparrow$ \\
\hline
Transformer & 0.8811 &  &  0.8036 &  & 0.7457  &  & 0.9073 &  & 0.9318 & \\
Transformer+FMNS & 0.8902 & 1.0\%$\uparrow$ & 0.8062 & 0.3\%$\uparrow$ & 0.7414 & 0.6\%$\downarrow$ & 0.9155 & 0.9\%$\uparrow$ & 0.9402 & 0.9\%$\uparrow$ \\
Transformer+FM & 0.9207 & 4.5\%$\uparrow\dagger$ & 0.8725  & 8.6\%$\uparrow\dagger$ & 0.7629 & 2.3\%$\uparrow$ & 0.9327 & 2.8\%$\uparrow$ & 0.9672 & 3.8\%$\uparrow\dagger$ \\
\hline
\end{tabular}
\end{table*}

\begin{table*}[t]
\centering\footnotesize
\caption{PC: performance of struggle detection (Up and down arrows indicate absolute performance increase and decrease. $\dagger$ shows statistically significant improvement from ``feature modulation (X+FM)'' runs over the original runs (one-tailed t-test, p$=$.01)). ``X+FMNS'' refers to ``Feature Modulation with No feature Selection''.} 
\label{tab-result-pc-2}
\begin{tabular}{|p{2.3cm}|p{0.7cm}|p{0.9cm}|p{0.7cm}|p{1cm}|p{0.7cm}|p{0.9cm}|p{0.7cm}|p{0.9cm}|p{0.7cm}|p{0.9cm}|}
\hline
 & accu. & impr. & pos. p & impr. & pos. r & impr. &  neg. p & impr. & neg. r & impr. \\
\hline
\hline
ZeroRule & 0.7079 & -- & -- & -- & 0.0000 & -- & 0.7079 & -- & 1.0000 & -- \\ 
\hline
\hline
LM & 0.8384 & & 0.7624 & & 0.8316 & & 0.8917 & &	0.8425 & \\
LM+FMNS & 0.8425 & 0.5\%$\uparrow$ & 0.7742  & 1.5\%$\uparrow$ & 0.8260 & 0.7\%$\downarrow$ & 0.8890 & 0.3\%$\downarrow$ & 0.8526 & 1.2\%$\uparrow$ \\
LM+FM &	0.8676 & 3.5\%$\uparrow\dagger$ &	0.8141 & 6.8\%$\uparrow\dagger$ &	0.8395 & 0.9\%$\uparrow$ &	0.9015 & 1.1\%$\uparrow$ &	0.8846 & 5.0\%$\uparrow\dagger$\\
\hline
SVM & 0.8617 &  &	0.7843 & &	0.8810 & &	0.9201 & &	0.8497 & \\
SVM+FMNS & 0.8757 & 1.6\%$\uparrow$ & 0.8008 & 2.1\%$\uparrow$ & 0.8810 & 0.0\% & 0.9265 & 0.7\%$\uparrow$ & 0.8726 & 2.7\%$\uparrow$ \\
SVM+FM & 0.9100 & 5.6\%$\uparrow\dagger$ &	0.8625 & 10.0\%$\uparrow\dagger$ &	0.8935 & 1.4\%$\uparrow$ & 0.9385 & 2.0\%$\uparrow$ & 0.9194 & 8.2\%$\uparrow\dagger$\\
\hline
\hline
Hassan et al.\cite{Hassan:2014:SED:2556195.2556221} & 0.8418 & &	0.7646 & &	0.8374 & & 0.8959 & & 0.8374 & \\
Hassan et al.+FMNS & 0.8604 & 2.2\%$\uparrow$ & 0.7927  & 3.7\%$\uparrow$ & 0.8416 & 0.5\%$\uparrow$ & 0.9040 & 0.9\%$\uparrow$ & 0.8714 & 3.2\%$\uparrow$ \\
Hassan et al.+FM &	0.8981 & 6.7\%$\uparrow\dagger$  &	0.8499 & 11.2\%$\uparrow\dagger$  &	0.8557 & 2.2\%$\uparrow$ & 0.9237 & 3.1\%$\uparrow\dagger$ & 0.9204 & 9.0\%$\uparrow\dagger$ \\
\hline
MART & 0.8775 & &	0.8223 & &	0.8605 & &	0.9134 & &	0.8878 & \\
MART+FMNS & 0.8952 & 2.0\%$\uparrow$ & 0.8416 & 2.3\%$\uparrow$ & 0.8683 & 0.9\%$\uparrow$ & 0.9262 & 1.4\%$\uparrow$ & 0.9100 & 2.5\%$\uparrow$ \\
MART+FM & 0.9293 & 5.9\%$\uparrow\dagger$ &	0.8874 & 7.9\%$\uparrow\dagger$ &	0.9024 & 4.9\%$\uparrow\dagger$ &	0.9508 & 4.1\%$\uparrow\dagger$ &	0.9428 & 6.2\%$\uparrow\dagger$\\
\hline
Transformer & 0.8813 & & 0.8266  & & 0.8649 &  & 0.9166 & & 0.8911 & \\
Transformer+FMNS & 0.9005 & 2.2\%$\uparrow$ & 0.8494 & 2.8\%$\uparrow$ & 0.8737 & 1.0\%$\uparrow$ & 0.9298 & 1.4\%$\uparrow$  & 0.9152 & 2.7\%$\uparrow$ \\
Transformer+FM & 0.9372 & 6.3\%$\uparrow\dagger$ & 0.8988 & 8.7\%$\uparrow\dagger$ & 0.9090 & 5.1\%$\uparrow\dagger$ & 0.9560 & 4.3\%$\uparrow\dagger$ & 0.9508 & 6.7\%$\uparrow\dagger$ \\
\hline
\end{tabular}
\end{table*}

\subsection{Experimental Setup}
\label{experimentsetup}
\subsubsection{Dataset Preparation} 

We collected a week's search log data from a commercial search engine in the period of 11/22/2020 $\sim$ 11/28/2020. We created two datasets, consisting of search logs using a PC browser and a mobile App. The user activities on the two platforms are slightly different due to different platform interfaces. 

We take the following steps to prepare our data. First, we segment the search log into topically coherent segments~\cite{10.1145/1458082.1458176}, each corresponding to a session. We segment the sessions following ~\cite{shuguang_cikm_16}. It uses logistic regression to classify two neighboring queries as they belong to the same search session or otherwise. Then, the consecutive query pairs are added into the same segment if they show high regression scores. The classification features include  query edits, click similarity and time-related features. We achieved a segmentation accuracy of 99.8\% for 10-fold cross-validation. 

Second, we recruit human assessors to annotate whether a session is struggling or non-struggling. 
The assessors were instructed to label a session into 1) Struggle, 2) Non-struggle, or 3) Uncertain. Each session was judged by two assessors independently. If there was a disagreement between the two assessors, a third assessor joined in resolving the dispute \cite{DBLP:conf/sigir/YangMSM10}. Every assessor carefully examined the query logs, with information about queries, user clicks, documents read by users, and timestamps of every user activity. The assessors went through a training session\footnote{The training materials and software interface can be made available upon acceptance.} before they started the actual annotation. Eventually, the annotations achieve inter-coder agreements of 73.3\%. Third, we removed ``Uncertain" sessions because the assessors could not determine if the session was likely to be struggling or not. 

In the end, we obtained 2,157 labeled sessions in total, including 601 struggling and 1,556 non-struggling sessions. We then processed these sessions to produce a feature vector for each of them. 
Table~\ref{tab-dataset_statistic} reports the dataset statistics. 

Fourth, we asked the assessors to mark out sessions that contained multiple search tasks. This step served as a sanity check for session segmentation.

\subsubsection{Baselines}

We experimented with several classifiers for struggle detection. They include common baselines as well as best-performing struggle detection methods. 

\begin{itemize} 
\item \textbf{ZeroRule} is a naive baseline that classifies  instances based on the majority label in the ground truth.

\item \textbf{SVM} is the support vector machine classifier~\cite{Cortes:1995:SN:218919.218929}, which is a top linear classifier. We use a radial kernel with a kernel coefficient of 0.016 and a cost of 2.0. 

\item \textbf{LM} is a logistic regression classifier \cite{friedman2000additive}, which is also a leading linear classifier.

\item \textbf{MART} is a Multiple Additive Regression Trees (MART) classifier~\cite{friedman2001greedy}, which is a top-performing non-linear classifier widely used in Web search. We set MART's n.tree to be 8000 and shrinkage 0.005.

\item \textbf{Transformer}~\cite{10.5555/3295222.3295349} is a binary classifier built on top of a multi-head self-attentive deep neural network. We use a batch size of 64, a learning rate of 0.00005, and a dropout rate of 0.1. All classifiers use features in Table \ref{tab-features} and the categorical feature, search topic. 

\item We also re-implement a state-of-the-art struggle detection system proposed by {\bf Hassan et al.}~\cite{Hassan:2014:SED:2556195.2556221} since it shares the most features with us. We used their features presented in \cite{Hassan:2014:SED:2556195.2556221} and experiment on our dataset. This model performs similarly compared to their reported results. All classifiers are trained and validated with 10-fold cross-validation.  
\end{itemize}

\subsection{Experimental Setup} For each baseline, we experiment with three different settings. (1) The original setting. (2) The baseline classifiers running with a variation of the proposed feature modulation method. We skip the ``means-ends" features step in the variation and directly use Eq. \ref{eq-modulation} to modulate all features. These runs  have suffix  ``+FMNS'', which stands for feature modulation no selection. (3) The baseline classifiers with only the ``means-ends" features are modulated. These runs have the suffix  ``+FM''. 

\subsection{Metrics} We evaluate the struggle detection systems using multiple metrics to understand their effectiveness from different perspectives. The metrics include {\it accuracy}, {\it positive precision} and {\it positive recall} (they are precision and recall for the  ``struggling" class), and {\it negative precision} and {\it negative recall} (they are precision and recall for the  ``non-struggling" class). Among them, {\it accuracy} and {\it positive precision} are the main metrics. Positive precision is important for web search~\cite{Savenkov:2014:HEE:2600428.2609523}. Because precise assistance is preferred over generic assistance by human users; hence precisely predicting user struggles is very important. 


 


\subsection{Struggle Detection Effectiveness}

\begin{figure}[t]
    \centering
    \begin{minipage}{.45\linewidth}
        \centering
       \includegraphics[width=\linewidth]{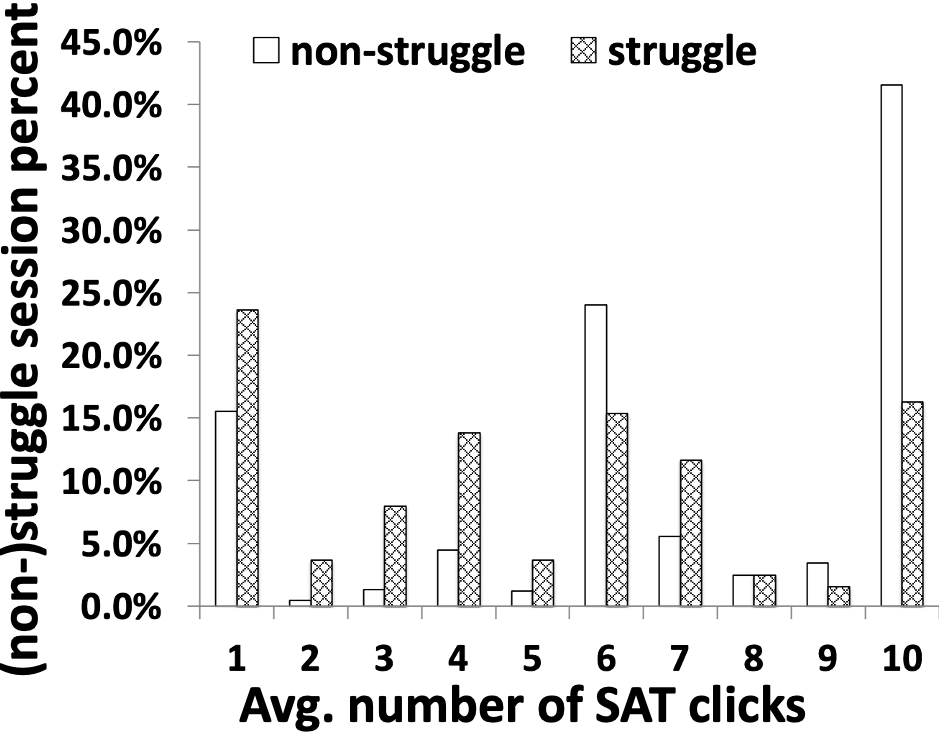} \text{before}
    \end{minipage}
    \begin{minipage}{0.45\linewidth}
        \centering
        \includegraphics[width=\linewidth]{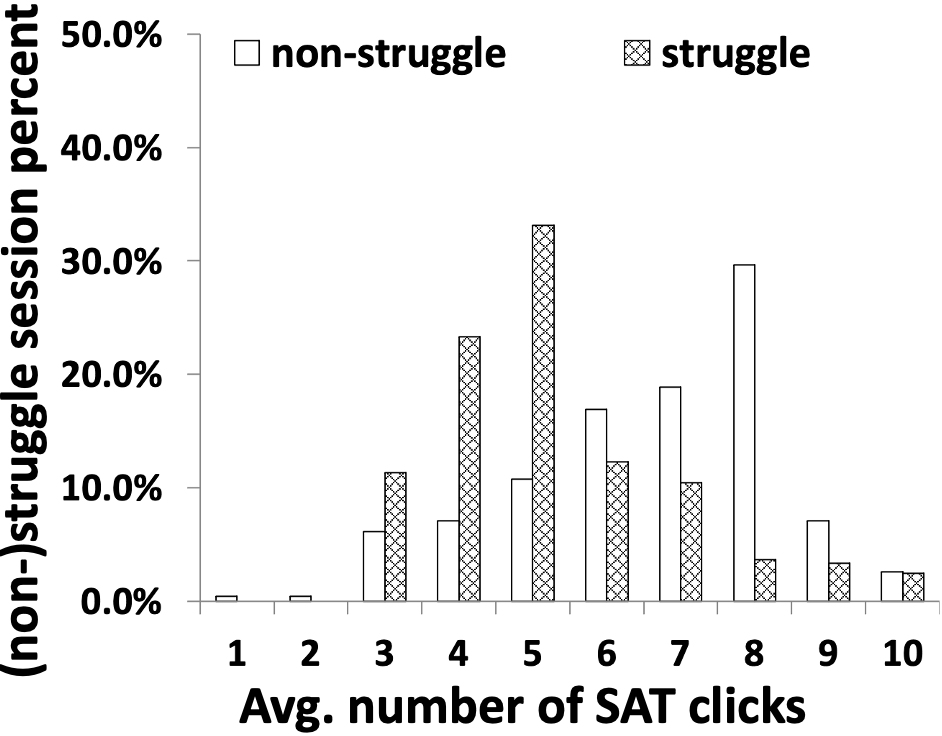} \text{after}
    \end{minipage}
    \caption{distribution of struggling and non-struggling sessions over avg. number of Satisfactory (SAT) clicks.} 
    \label{fig-avg-sat-click}
\end{figure}

\begin{figure}[t]
    \centering
    \begin{minipage}{.45\linewidth}
        \centering
       \includegraphics[width=\linewidth]{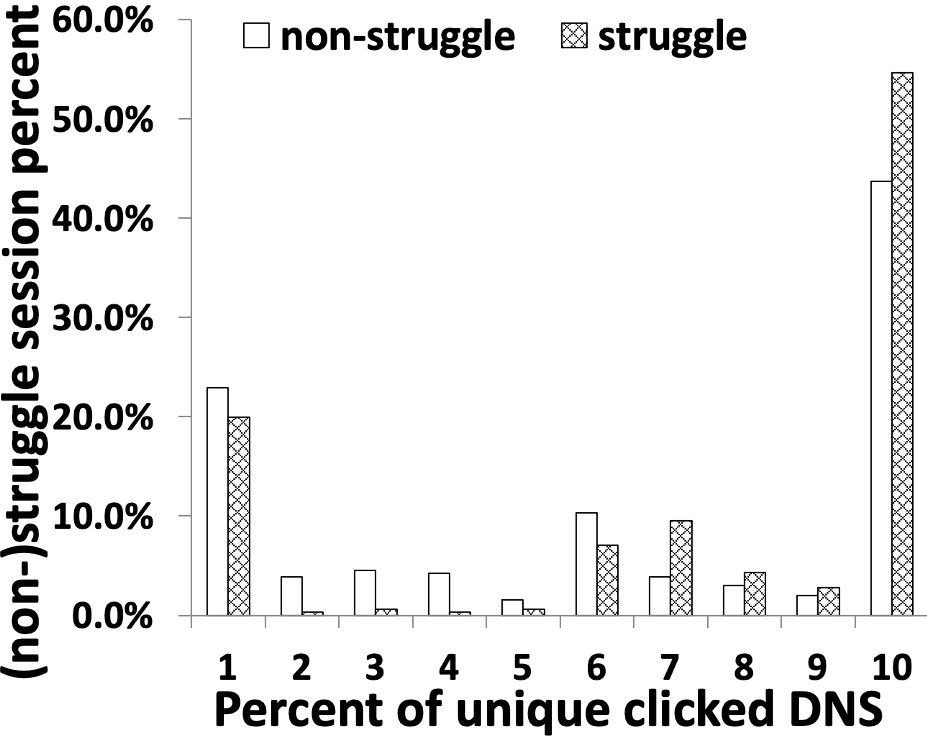}  \text{before}
    \end{minipage}
    \begin{minipage}{0.45\linewidth}
        \centering
        \includegraphics[width=\linewidth]{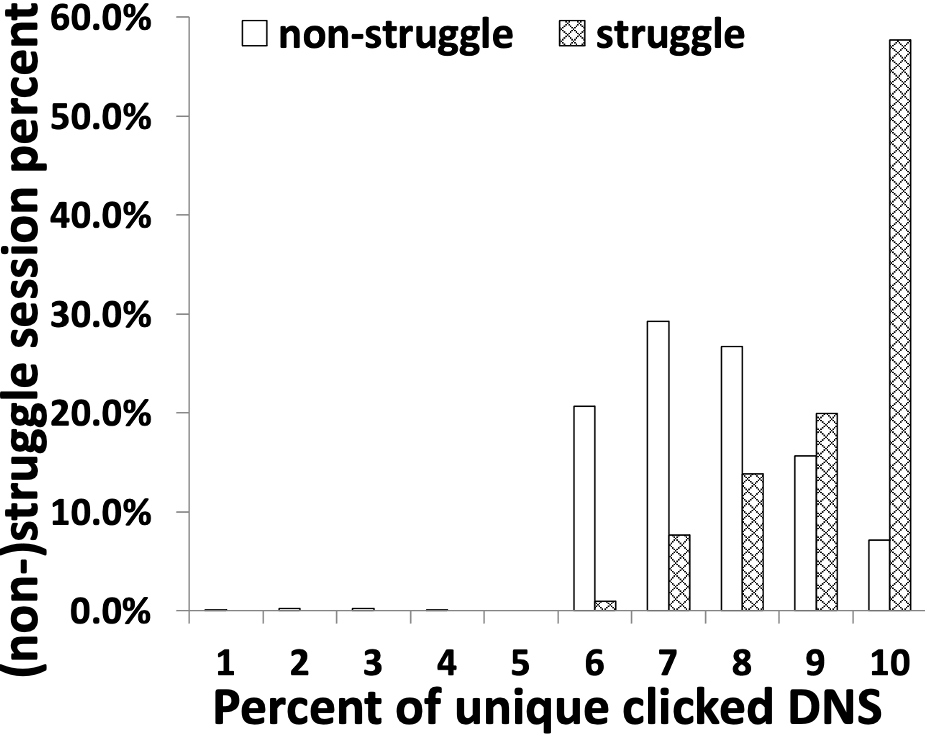}  \text{after}
    \end{minipage}
    \caption{distribution of struggling and non-struggling sessions over the percentage of unique clicked DNS domains.} 
    \label{fig-unidomain}
\end{figure}

Tables \ref{tab-result-1} and \ref{tab-result-pc-2} report the effectiveness of struggle detection by the runs compared to the mobile and PC datasets. 
We also highlight the percentage improvement of a run from its original run. In addition, we report  one-tailed t-test statistical results between the ``+FM'' runs and the initial runs.

The results show that the proposed feature modulation method is highly effective. The ``+FM'' runs statistically significantly improve the performance of all classifiers on all metrics. On average, our approach boosts a baseline method's accuracy by $\sim$5\% and  positive precision by $\sim$9\%. 
Combined with our method, these classifiers have become highly effective. Transformer+FM achieves the best performance among all models and settings, with a high 0.937 accuracy and 0.899 positive precision for the PC dataset. We observe similar trends on the mobile dataset. The ``+FMNS'' runs gain slightly better performance than the original baselines and worse than the "+FM" runs. 
It confirms what RT suggests that only the first dimension, ``means-ends", impacts the arousal model, thus effective on our struggle detection task. Other features, some of which are more related  to the ``rules" dimension, which RT considers irrelevant. The weak performance from the ``+FMNS'' runs again supports this insight from RT, besides our hypothesis test in Section \ref{sub_sec_rule_dimension}.  

\subsection{Impact of Feature Modulation}

We further investigate the effect on individual features before and after feature modulation. In this investigation, we study the magnitude of the features for both struggling and non-struggling sessions and their distributions. Figures \ref{fig-avg-sat-click} and \ref{fig-unidomain}  plot the effect of modulating two features, avg. number of SAT clicks and percentage of unique clicked DNS domains, respectively. 
The figures are generated by first dividing the magnitudes of avg. SAT clicks per session and percentage of unique clicked domains into ten bins evenly and then plotting the ratio of struggling and non-struggling sessions in each bin. As we can see, before feature modulation, the distributions of struggling and non-struggling sessions do not present any exciting patterns. After feature modulation, however, the distributions of struggling and non-struggling sessions are centered at different bins. It suggests our method  helps the two features better separate the binary classes and become more valuable features.

\section{Conclusion}




This paper takes a unique solution path, uses insights from established psychology theories, and derives working solutions from there. 
Inspired by the Reversal Theory in psychology, we propose a novel feature modulation method to work as a component in Web search struggle detection. Our work is an initial application of the reversal theory, which we think will show great potential in other sub-fields of information retrieval. For instance, users' psychological states can be reversed when proper triggers present. To reverse from telic to paratelic state, situational triggers can be removal of threat, entertainment, or humor. Interestingly enough, Web searcher struggle can be a trigger to reversal. It implies that across multiple sessions, if a struggling session is detected, we may take advantage of this knowledge to infer states for the later sessions. We leave these interesting directions as future work.

\bibliographystyle{abbrv}
\bibliography{bibs/cikm,bibs/kdd,bibs/wsdm,bibs/www,bibs/chi,bibs/journal,bibs/sigir,bibs/minor,bibs/evaluation}  

\balance

\end{document}